\documentclass[aps, prd, preprint, a4paper,12pt]{revtex4-2}
\usepackage[utf8]{inputenc}
\usepackage{amsmath}
\usepackage{hyperref}
\usepackage{amsfonts}
\usepackage{amssymb}
\usepackage{graphicx}
\usepackage{latexsym}
\usepackage{color}
\usepackage{booktabs}
\usepackage{dcolumn}
\usepackage{epsfig}
\usepackage{subfigure}
\usepackage{float}
\usepackage{multirow}
\usepackage{ulem}
\usepackage{xcolor} 
\usepackage{color}
\usepackage{svg}
\usepackage{tikz}
\usepackage[mathcal]{euscript}
\def\be{\begin{eqnarray}}
\def\ee{\end{eqnarray}}

\begin{document}

\title{Aspects of regular and singular electromagnetic-generalized-quasitopological-gravities black holes in $2+1$ dimensions}

\author{Jeferson de Oliveira}\email{jeferson.oliveira@ufmt.br}
\affiliation{Instituto de F\'i­sica, Universidade Federal de Mato Grosso, Cuiab\'a, Mato Grosso 78060-900, Brazil}

\author{R. D. B. Fontana}
\email{rodrigo.dalbosco@ufrgs.br}
\affiliation{Universidade Federal do Rio Grande do Sul, Campus Tramandaí-RS
\\
Estrada Tramandaí-Osório, Rio Grande do Sul 95590-000, Brazil}
\affiliation{Departamento de Matemática, Universidade de Aveiro}
\affiliation{Center for Research and Development in Mathematics and Applications (CIDMA), Campus Santiago, 3810-283, Aveiro, Portugal}

\author{A. B. Pavan}
\email{alan@unifei.edu.br}
\affiliation{Instituto de F\'i­sica e Qu\'i­mica, Universidade Federal de Itajub\'a, Itajub\'a, Minas Gerais 37500-903, Brazil}

\date{\today}

\begin{abstract}

We investigate quasitopological black holes in $(2+1)$ dimensions in the context of electromagnetic-generalized-quasitopological-gravities (EM-GQT). For three different families of geometries of quasitopological nature, we study the causal structure and their response to a probe scalar field. To linear order, we verify that the scalar field evolves stably, decaying in different towers of quasinormal modes. The studied black holes are either charged geometries (regular and singular) or a regular Ba\~nados-Teitelboim-Zanelli (BTZ)-like black hole, both coming from the EM-GQT theory characterized by nonminimal coupling parameters between gravity and a background scalar field. We calculate the quasinormal modes applying different numerical methods with convergent results between them. The oscillations demonstrate a very peculiar structure for charged black holes: in the intermediate and near extremal cases, a particular scaling arises, similar to that of the rotating BTZ geometry, with the modes being proportional to the distance between horizons. For the single horizon black hole solution, we identify the presence of different quasinormal families by analyzing the features of that spectrum. In all three considered geometries, no instabilities were found.

\end{abstract}

\maketitle
\newpage


\section{Introduction}

Black hole solutions in lower dimensional gravity can be a good testing ground for four-dimensional gravity ideas since they are simpler and share various features with the black hole solutions in higher dimensions. Aspects of gravity solutions in two and three dimensions with flat and anti-de Sitter/de Sitter (AdS/dS) asymptotic spatial infinity were extensively studied in the seminal work of Jackiw \cite{Jackiw:1984je}(for a detailed review on black hole solutions in lower dimensions, see also \cite{Mann:1991qp}). Another pioneer breakthrough was the Banãdos-Teitelboim-Zanelli (BTZ) black hole \cite{Banados:1992wn}, a three-dimensional solution whose spatial infinity is AdS-like having well-defined conserved charges as mass, angular momentum, and electric charge. Hennigar {\it {et al.}} \cite{Hennigar:2020fkv} found a three-dimensional black hole solution with Gauss-Bonnet term nonminimally coupled to a scalar field, which recovers the BTZ black hole when the Gauss-Bonnet coupling constant goes to zero. BTZ black holes with higher curvature corrections were considered in \cite{Konoplya:2020ibi}. A family of black hole solutions in three dimensions exhibiting Lifshitz scaling in the new massive gravity was considered in \cite{Ayon-Beato:2009rgu}. For a comprehensive analysis, identification, and algebraic categorization of exact solutions in $(2+1)$-dimensional Einstein gravity,  refer to \cite{Garcia-Diaz:2017cpv}. It provides valuable insights on low-energy $(2+1)$-dimensional string gravity, black holes coupled to nonlinear electrodynamics, and a general discussion on Einstein equations coupled to matter and fields.

More recently, Bueno {\it{et al.}} \cite{Bueno:2021krl} found another family of three-dimensional black hole solutions in the context of the so-called electromagnetic-generalized-quasitopological-gravities (EM-GQT). Such gravity theory is shaped considering a specific lagrangian density that encodes the Einstein-Hilbert term plus a nonminimal coupling between kinetic terms of the real scalar field $\partial_{a}\phi$, the Ricci scalar $R$ and Ricci tensor $R_{ab}$, respectively.
The black hole solutions are obtained through the imposition of what was called a "magnetic" ansatz to the scalar field i. e., $\phi$ proportional to the compact angular coordinate $\varphi$, and the metric tensor components obeying $g_{tt}g_{rr} = -1$. Depending on the choice of the undetermined coupling constants of the lagrangian density, there are black hole solutions with one or multiple horizons and with or without singularities, representing a generalization of the BTZ black holes \cite{Banados:1992wn}. The class of solutions describing regular EM-GQT black holes is exquisite as long as the formation of singularities in lower-dimensional gravity and general relativity (GR) reveals the breakdown of such theories requiring the inclusion of quantum corrections \cite{Penrose:1964wq, Hawking:1967ju}.


In $(2+1)$ dimensions, the theory proposed by \cite{Bueno:2021krl} is based on their counterparts in higher dimensions \cite{cano2020resolution, Cano_2020}. Typically, theories with higher order curvature terms extend the understanding of a zeroth order approach of curvature such as GR in four dimensions or that considered by Bañados {\it et al.} in \cite{Banados:1992wn}. In \cite{cano2020resolution, Cano_2020}, a general electromagnetic term nonminimally coupled to curvature is schematically described as yielding regular conditions for the gravitational and electromagnetic fields, namely singularity-free black holes. Beyond that benefit of presenting regular spacetimes with horizons as black hole mimickers (also know as regular black holes), the study of AdS geometries in such comes with nontrivial properties of their conformal field theory (CFT) counterparts where the spacetime molded peculiar behavior as superconductivity is to be unveiled.

The lower-dimensional theory of \cite{Bueno:2021krl} represents an extension of the traditional $(2+1)$ gravity with a very interesting specificity not to be seen in the higher dimensional counterpart: the duality of the electromagnetic source to a coupled scalar field theory system. In particular, whenever such a scalar field is given in terms of an azimuthal one-form (magnetic gauge), the theory presents a very elegant solution [see (\ref{func_horizonte}], below) with a multitude of possible black holes.


Different models characterizing and building regular black holes have been considered since the work of Bardeen \cite{bardeen1968non}. Exact regular solutions with the inclusion of nonlinear electromagnetic fields were studied in \cite{Ayon-Beato:1998hmi}. Gravity nonminimally coupled to Maxwell electromagnetic fields were considered in \cite{Balakin:2007am}.

This work considers three significant cases of EM-GQT black holes: a singular solution with Cauchy and event horizons and two regular solutions with one and two horizons. Our aim is twofold: the characterization of the geometry of those EM-GQT three-dimensional black holes to understand the physics in inner and outer regions; we also address the question of spacetime stability doing the computation of scalar field perturbations and quasinormal modes (QNM) ignited by a probe scalar field. Since the ringdown phase of perturbed black holes is characterized by the quasinormal modes \cite{Berti:2009kk, Berti:2007dg} and independent of the initial perturbation, the QNM can bring important information about the properties of black hole geometry. The oscillations represent part of the observations in the gravitational interferometers \cite{LIGOScientific:2016aoc, LIGOScientific:2016sjg, LIGOScientific:2017bnn}. The QNM can also shed light on the question of the linear stability of black hole solution under small perturbations modeled by classical probe fields \cite{Abdalla:2019irr, Fontana_2019, Abdalla:2018ggo, FernandezPiedra:2010wvn, Abdalla:2012si, Fontana:2022whx, Destounis:2020yav}. Additionally, QNM plays an essential role in the context of AdS/CFT conjecture \cite{Birmingham_2002}, in which the relaxation timescale for the dual field theory at finite temperature is given by the inverse of the imaginary part of the fundamental quasinormal mode \cite{Horowitz_2000, Abdalla:2011fd, Pan:2009xa, Lin:2014vla, deOliveira:2018weu}.

In (2+1) dimensions, the quasinormal modes were richly studied in the past. They were first calculated in \cite{Cardoso_2001} and \cite{0101194} providing a deep understanding of the AdS/CFT duals \cite{0305113} and scrutinized in several different works \cite{1906.06654, 1906.04360, 1904.10847, 1901.00448, 1801.02555, 1801.03248, 1711.04146, 1404.5371, 1407.6394, 1003.1381, 0908.0057, 0903.1537,0903.0088,0802.3321, 0306214}. Here, we will analyze the structure of the different quasinormal spectra in light of the studies performed in the past and their developments.

Our paper is organized as follows. In Sec. \ref{sec.2}, we review the essential aspects of the EM-GQT black hole solutions and explore the conformal structure. In Sec. \ref{sec.3}, we discuss the methods for finding the QNM due to the propagation of a massless probe scalar field. Finally, Sec. \ref{sec.final} summarizes the results and discusses possible open questions.

\section{Black hole solutions}\label{sec.2}

In this paper, we are going to consider black hole solutions of the equations of motion coming from the action \cite{Bueno:2021krl}
\begin{equation}\label{action}
S = \frac{1}{16\pi G}\int \sqrt{-g}d^{3}x\left[R+\frac{2}{L^{2}}-\Gamma \right],
\end{equation}
where $G$ is the gravitational constant in $(2+1)$ dimensions, $L$ is the AdS-like curvature radius, $R$ is the Ricci scalar and $\Gamma$ contains  terms concerning to the coupling between curvature and the scalar field $\phi$,
\begin{equation}\label{gamma}
\Gamma = \sum_{n=1}\alpha_{n}L^{2(n-1)}\left(\partial\phi\right)^{2n} - \sum_{m=0}\beta_{m}L^{2(m+1)}\left(\partial\phi\right)^{2m}\left[(2m+3)R^{ab} - g^{ab}R\right]\partial_{a}\phi\partial_{b}\phi.
\end{equation}
The dimensionless constants $\alpha_{n}$ and $\beta_n$ are entirely arbitrary, and their physics depend on the specific solution of the theory (\ref{action}).


The action expressed in (\ref{action}) brings $\Gamma$ as a quasitopological (not invariant) term similar to that of other quasitopological gravities (QTGs) studied along the past decades \cite{Myers_2010}. In such theories, gravitation comes as a consequence of an action with all possible independent cubic curvature contractions of the Riemann tensor. The invariant brings interesting metric solutions with AdS ansatz scrutinized in different studies and contrarily to the similar topological theories as e. g. that of Lovelock gravities. In the QTGs, we do not have $S$ as an invariant of the spacetime and the border acts nontrivially delimiting the geometry, depending on the chosen topology for the spacetime \cite{Myers_2010, cano2020resolution, Cano_2020}.


We will consider the following ansatz for the line element $ds^{2}$ and the background scalar field $\phi$,
\begin{equation}\label{ansatz}
ds^{2} = -f(r)dt^{2}+f(r)^{-1}dr^{2} +r^{2}d\varphi^{2}, \hspace{0.3cm} \phi = p\varphi,
\end{equation}
with $p$ standing for an arbitrary dimensionless constat (that can be real or purely imaginary) and $(t,r,\varphi)$ denoting the coordinate system in which $t\in[0,+\infty[$, $r\in[0,+\infty[$ and $\varphi\in[0,2\pi]$. In this case, setting $\alpha_1=0$ the solution of the motion equations is given by
\begin{equation}\label{func_horizonte}
f(r) = \left(\frac{r^{2}}{L^{2}}-\mu+\sum_{n=2}\frac{A_{n}}{r^{2(n-1)}}\right)\left(1+\sum_{m=0}\frac{B_{m}}{r^{2(m+1)}}\right)^{-1},
\end{equation}
where $A_{n} = \alpha_{n}p^{2n}L^{2(n-1)}/2(n-1) $ and $B_{m} = (2m+1)\beta_{m}(L p)^{2(m+1)}$ and $\mu$ is an integration constant which is written in terms of the black hole mass $M$ as
\begin{equation}
M=\mu +\beta_{0}p^{2} + \alpha_1 p^{2}\log{(\frac{r_{0}}{L})},
\end{equation}
where $r_{0}$ is a cutoff radius. Such a mass $M$ is divergent as $r_{0}\rightarrow \infty$ as in the charged BTZ solution \cite{Martinez:1999qi}. For the black hole solutions considered in the present work, we have set $\alpha_1=0$, so the cutoff is unnecessary. If all $A_{n}$ and $B_{m}$ coefficients are zero, the spacetime corresponds to the simplest BTZ black hole \cite{Banados:1992wn}. 


The lowest terms in (\ref{func_horizonte}), namely $\alpha_{1,2}$ and $\beta_1$ provide us regular (and singular) black holes solutions representing a nice first order theory: e. g. the lapse function of (\ref{bh_singular}) and (\ref{regular1_dois_horizontes}) resembles that of Reissner-Nordstr\"om solutions in GR and in such (as we report in our results) possesses a family of oscillations proportional to the distance between horizons. However, unlike the four-dimensional black holes, where such a family of oscillations is subdominant \cite{Cardoso_2018, papa21}, in the particular solutions we treat, they play the commanding role in the quasinormal spectrum.

If we consider (\ref{gamma}) in a context of small coupling factor (lower alphas and betas) the charged nonrotating BTZ black hole ($\alpha_1 \neq 0$ and $\alpha_{\neq 1} = \beta_n =0$) is a possible solution of (\ref{func_horizonte}) studied in \cite{f23quasi} as also the charged BTZ black holes consequently from nonlinear electrodynamical terms (see \cite{aragon_2021} and references therein). As expected, whenever the extra constants are turned off, (\ref{func_horizonte}) recovers the Bañados-Teitelbonn-Zanelli geometry and hence we can consider it as a deformation of a zeroth order theory.


The line element (\ref{ansatz}) together with (\ref{func_horizonte}) describes a family of static and circular symmetric black holes with or without curvature singularity, depending on how many constants $\alpha_{n}$ and $\beta_{m}$ are turned on in the theory.


As mentioned, the magnetic ansatz for the scalar field \cite{Bueno:2021krl} corresponds to make it proportional to the form $\varphi$. In the EM-GQT theory, the density $\Gamma$ is dual to an electromagnetic Lagrangian
\be
\label{ed1}
\mathcal{L}_{dual} = R+\frac{2}{L^2}-\Gamma - F_{ab}\varepsilon^{abc} \partial_c \phi
\ee
in which the electromagnetic tensor $F$ is written as
\be
\label{ed2}
F_{ab} = -\frac{1}{2}\varepsilon_{abc} \frac{\partial \mathcal{L}}{\partial \partial_c \phi}.
\ee
When solved perturbatively, such ansatz allows for a relation between $\phi$ and $F$ reducing the latter to its typical form, of known usual charged black holes, that of an electromagnetic tensor $F$ produced solely by a source term $A$ of type $A= A_adx^a = A_t(r) dt$. In that case, the solution for the source $A$ is of electric type (as usually in charged black holes) written depending on the coefficients of the theory as
\begin{equation}\label{ed3}
A_{t}(r) = -\alpha_1 p \log{(r/L)} + \sum_{n=2}\frac{n\alpha_{n}p}{2(n-1)}\left(\frac{L p}{r}\right)^{2(n-1)} + f'(r)L\sum_{m=0}\beta_{m}(m+1)\left(\frac{L p}{r}\right)^{2m+1}.
\end{equation}

We will analyze three different cases in what follows, considering the presence of only one extra (geometric) parameter besides those of a BTZ black hole with mass and negative cosmological constant.

The first family of black holes considered is singular, featuring event and Cauchy horizons, with a function $f(r)$ identical to the four-dimensional Reissner-Nordstr\"om-AdS black hole a very interesting peculiarity of the solution.

The other two families of black holes are regular; each represents a different type of regular solution with an event horizon in EM-GQT. The first type corresponds to regular solutions in which $f(r=0)=1$ and a constrained parameter $p=\sqrt{2\beta_{0}/\alpha_{2}}$ [see Eq. (\ref{regular1_dois_horizontes})]. For the second type, $f(r\rightarrow0)\rightarrow\mathcal{O}(r^{2s})$, with $s\ge 1$ [see Eq. (\ref{regular2_um_horizonte})]. In both regular cases, the curvature invariants are finite everywhere.

\subsection{Singular black hole with two horizons: SBH }
Let us consider the solution with one constant $\alpha_2$ and set $\alpha_{1,n\ge 3}=\beta_{m}=0$. Then the line-element results in a singular black hole,
\begin{equation}\label{bh_singular}
ds^{2} = -\frac{r^{2}}{L^{2}}\left(1-\frac{(r_{+}^{2}+r_{-}^{2})}{r^{2}}+\frac{(r_{+}r_{-})^{2}}{r^{4}}\right)dt^{2}+\frac{L^{2}dr^{2}}{r^{2}\left(1-\frac{(r_{+}^{2}+r_{-}^{2})}{r^{2}}+\frac{(r_{+}r_{-})^{2}}{r^{4}}\right)}+r^{2}d\varphi^{2},
\end{equation}
featuring two solutions $f(r)=0$, the event and Cauchy horizons, respectively, at $r =r_{+}$ and $r=r_{-}$, or,
\begin{equation}\label{horizontes}
r_{\pm} = L\sqrt{\frac{\mu}{2}}\left[1\pm\left(1-\frac{2\alpha_2 p^4}{\mu^2}\right)^{1/2}\right]^{1/2},
\end{equation}
with surface gravity at the horizons $k_{\pm} = |r_{+}^{2}-r_{-}^{2}|/L^{2}r_{\pm}$. Notice that in the extremal case in which $r_{+}=r_{-}$ the surface gravity goes to zero similarly to the four-dimensional extremal Reissner-Nordstr\"om black hole. Apart from such occasion, the black hole given by (\ref{bh_singular}) has a timelike singularity at the origin covered by the event and Cauchy horizons. The spatial infinity is conformally AdS-like as in other BTZ-like geometries.  

The curvature scalar $K=R^{abcd}R_{abcd}$ \footnote{$R_{abcd}$ are the components of Riemann tensor.} reads
\begin{equation}\label{kresh_singular_2hor}
K = \frac{12}{L^4} +8\left(\frac{r_{-} r_{+}}{L^{2}}\right)^{2}\frac{1}{r^{4}} + 44\left(\frac{ r_{-} r_{+}}{L}\right)^{4}\frac{1}{r^{8}},
\end{equation}
bringing a curvature singularity at $r=0$ and positive constant value $K\sim \frac{12}{L^4}$ at $r\to\infty$. For a more detailed description of the nature of the singularity we proceed to the Penrose-Carter diagram.

\begin{figure}[htb]
\begin{center}
\includegraphics[height=11cm, width=10cm]{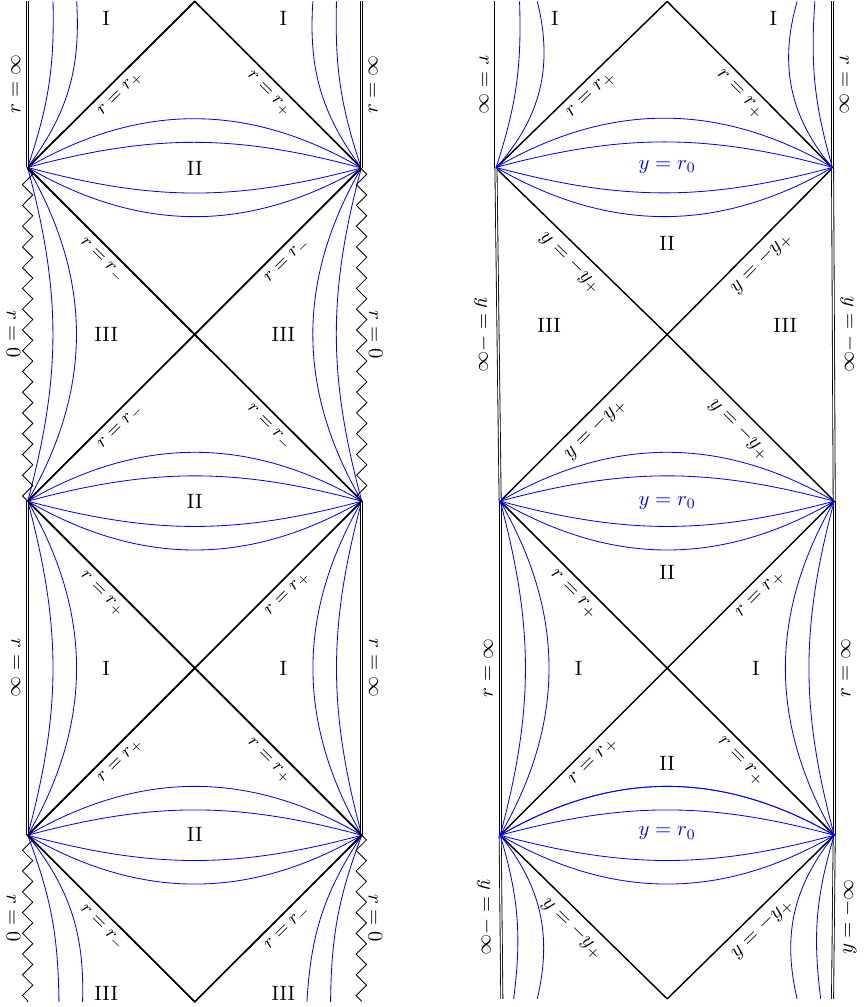}
\caption{{\it{Left:}} Conformal diagram for maximal extended singular black hole solution with two horizons $\alpha_2\neq 0$. {\it{Right:}}  Conformal diagram for regular black hole with one event horizon.The blue lines represent the surfaces in which the radial coordinate is constant.}
\label{diagrams}
\end{center}
\end{figure}

The  Kruskal-Szekers-like coordinates $(U,V)$ adapted to the event horizon $r_{+}$ are
\begin{equation}\label{ks-event}
U_{+}V_{+} = \mp \left|\frac{r-r_{+}}{r+r_{+}}\right|\left|\frac{r+r_{+}}{r-r_{-}}\right|^{k_{+}/k_{-}},
\end{equation}
where the upper sign refers to $r>r_{+}$ and the lower sign refers to $r<r_{-}$. For the Cauchy horizon $r_{-}$, they have a new set of coordinates $(U,V)$
\begin{equation}\label{ks-cauchy}
U_{-}V_{-} = \pm \left|\frac{r-r_{-}}{r+r_{-}}\right|\left|\frac{r+r_{+}}{r-r_{+}}\right|^{k_{-}/k_{+}},
\end{equation}
in which the upper sign refers to the region $r_{-} \le r \le r_{+}$ and the lower sign refers to $0\le r \le r_{+} $.
These relations allow us to perform a compactification of the spacetime regions and then combine the overlapping coordinates patches $(U_{+}, V_{+})$ and $(U_{-}, V_{-})$ to obtain the conformal diagram of (\ref{bh_singular}). That is shown in Fig. \ref{diagrams} (left panel). 

The diagram structure is identical to the four-dimensional Reissner-N\"ordstrom-AdS black hole \cite{Griffiths:2009dfa}. Notice that the whole spacetime consists of an infinite lattice of identical blocks, each containing an event and Cauchy horizon and a timelike singularity at $r=0$. Due to the repulsive character of the timelike singularity, a real particle in region III will cross $r=r_{-}$ and emerge in a white hole horizon with asymptotically AdS-like outer region I.

As explored in \cite{Bueno:2021krl} the dual frame description of the theory (\ref{action})\footnote{The action in dual description can only be obtained perturbatively.} allows to consider the constant $q_{e} =\alpha_2 p^{3}L^{2}$ as an ``electric charge" with the  electrostatic potential $ A_{t} = q_{e} r^{-2}$. The particular shape of the source $A$ in this case is the same as that of the non-linear electrodynamics theory (Einstein-Power-Maxwell) \cite{Gurtug_2012,pan19} if we consider a density of type $\mathcal{L}=(F_{\mu \nu}F^{\mu \nu})^{2/3}$. In such theory, $q_e$ can be interpreted as the electrical charge of the non-linear density $\mathcal{L}$ and the source term corresponds exactly to that of our gravity with only $\alpha_2 \neq 0$ thus justifying the interpretation of that quantity as electric charge.

Such a result agrees with the conclusion that the effective potential of an uncharged real particle in region  III near the timelike singularity is repulsive, preventing the particle's motion from terminating at $r=0$.
\subsection{Regular black hole with two horizons: RBH}

Considering $f(r)$ in (\ref{func_horizonte}) with $\alpha_2\neq 0$ and $\beta_0 = \frac{\alpha_2 p^2}{2} $, we have the solution describing a regular black hole solution endowed with event and Cauchy horizons $r_{+}$ and $r_{-}$, respectively,
\begin{equation}\label{regular1_dois_horizontes}
f(r) = \frac{r^2}{L^2}\left(1-\frac{(r_{+}^2+r_{-}^{2})}{r^2} +\frac{(r_{+}r_{-})^{2}}{r^{4}}\right)\left(1+\frac{(r_{+}r_{-})^{2}}{L^2r^2}\right)^{-1},
\end{equation}
with
\begin{equation}\label{horizons_regular1}
r_{\pm} = L\sqrt{\frac{\mu}{2}}\left[1\pm\left[1-\frac{2 \alpha_2 p^4}{\mu^{2}}\right]^{1/2}\right]^{1/2}.
\end{equation}
Notice that the event and Cauchy horizons locations are the same as in the case of the singular black hole. It is convenient to define the parameter $Q=\sqrt{r_{-}r_{+}/L}$, associated to the presence of the background scalar field that we will use as a parameter of the black hole extremality ($Q_{ext}=r_+$) in the numerical calculations presented in next section. We emphasize that such parameter is not associated with a charge defined via source term $A$, differently from the previous case.

The manifest difference with the first spacetime is the absence of a curvature singularity at $r = 0$ in the context of solution (\ref{regular1_dois_horizontes}). In such a case, $r=0$ is a well-behaved surface that is confirmed by the curvature scalar $K$ when $r\to 0$ and written as
\begin{equation}\label{kresh_regular_2hor}
K \approx \frac{12 \left(L^2+r_{+}^2+r_{-}^2\right)^2}{\left(r_{+} r_{-}\right)^{4}} + O\left(r^2\right).
\end{equation}
Near to the spatial infinity $(r \to \infty)$ the curvature scalar has the approximate value of BTZ black hole curvature $K\sim K_{BTZ}=\frac{12}{L^4}$.   

The regular black hole with two horizons has the same Kruskal-Szekes extension as the singular black hole (\ref{ks-event}), (\ref{ks-cauchy}). But now, being $r=0$ a regular surface, the manifold extends beyond it. The surface gravity at event and Cauchy horizons are, respectively
\begin{eqnarray}
  k_{+} = \frac{|r_{+}^{2} - r_{-}^{2}|}{r_{+}\left(r_{-}^{2}+L^{2}\right)},\\
  k_{-} = \frac{|r_{+}^{2} - r_{-}^{2}|}{r_{-}\left(r_{+}^{2}+L^{2}\right)}.
\end{eqnarray}
The Penrose-Carter diagram for the regular black hole with two horizons has the same structure as in the previous case for the singular solution (left diagram in Fig. \ref{diagrams}) except for the timelike singularity which is absent in the regular solution.

On a timelike trajectory, a particle with zero angular momentum can go through the regular surface $r=0$, since the effective potential is finite in this region, otherwise the potential is infinitely repulsive preventing the particle to cross $r=0$, similarly to the singular black hole solution (\ref{bh_singular}) and the four-dimensional Reissner-Nordstr\"om AdS black hole \cite{Griffiths:2009dfa}.

\subsection{Regular black hole with one horizon and without charge}

Another representative regular black hole solution is obtained by considering $\alpha_{n} = 0$, $\beta_{m>0}=0$  and only $\beta_0\neq 0$. In this case (\ref{func_horizonte}) reduces to
\begin{equation}\label{regular2_um_horizonte}
f(r) = \frac{r^{2}}{L^{2}}\left(1-\frac{r_{+}^{2}}{r^{2}}\right)\left(1+\frac{\beta_{0}p^{2}L^{2}}{r^{2}}\right)^{-1},
\end{equation}
in which we choose $p$ to be real and $\beta_0 >0 $\footnote{The choice of $p$ purely imaginary or equivalently $\beta_0 <0$ would bring a singular spacetime with a singularity at $r=\lambda$ that we will not study here.}.
The spacetime is asymptotically AdS-like and has an event horizon at $r=r_{+}$ with a regular origin at $r=0$. In Fig. \ref{diagrams} (right panel), we performed a change in the radial coordinate $r = r_{0}+y$, where $r_{0}$ is a positive constant giving the size of the throat of the black hole and $y\in ]-\infty,+\infty[$. For a timelike geodesic entering region II of the right diagram, the effective potential in the vicinity of $r=0$ is finite allowing massive particles to cross the region $y=r_{0}$ reaching region III.

The curvature scalar $K$ when $r\to0$ is given by
\begin{equation}\label{kresh_regular_1hor_lambda}
K \approx \frac{12 r_{+}^4}{\beta_{0}^2p^{4} L^8}-\frac{80 \left(\beta_{0}L^{2}p^{2}r_{+}^2+r_{+}^4\right)}{\beta_{0}^3 p^6 L^{10}}\ r^2 + O\left(r^3\right),
\end{equation}
being regular at the origin as aforementioned. In this case, near to the spatial infinity $(r \to \infty)$ the curvature scalar has the approximate value of BTZ black hole curvature $K\sim K_{BTZ}=\frac{12}{L^4}$, too.

In the previous analysis, from the three representative black hole solutions studied, two of them are regular (\ref{regular1_dois_horizontes}),(\ref{regular2_um_horizonte}) and another has a timelike singularity at the origin (\ref{bh_singular}). The regular solution without charge is the most simple and the first (to the best of our known) regular black hole in (2+1) dimensions similar to that of the BTZ geometry endowed with an extra scalar hair $\lambda$
\be
\label{eqe4}
\lambda = \sqrt{\beta_0} p L 
\ee
such that the lapse function is defined in terms of three parameters $L$, $r_+$ and $\lambda$. By inspection we may verify that such parameter is directly defined by the coupling constant of the theory and the other two spacetime properties, $r_+$ and $L$ as long as in this case,
\be
\label{eqe5}
\Gamma = -8\lambda^4\frac{(r_+^2+L^2)}{L^2(r^2+\lambda^2)^2}.
\ee
When $\lambda$ is turned off, we recover the pure BTZ spacetime.

We recall that in the asymptotic spatial infinity region, the curvature of all three spacetimes is constant and same as that of the BTZ spacetime curvature.  

In the lapse function (\ref{regular2_um_horizonte}) the redefinition $\lambda = \beta_0^{1/2}p L$ allows us the reabsorption of the coupling constant $p$ of the electric ansatz (or the scalar one in the dual scheme) in terms of the coupling constants $\alpha$´s and $\beta$´s  of the theory or vice-versa.

That is also the case for the other two black holes considered above as it can be seen trough the potential $A(r)$. In this work we consider solutions of such theory with only one extra geometric parameter, $\lambda$ or $r_-$ that generate the above black holes regular and singular with multiple or single horizons. In this fashion, we studied the influence that this extra property can produce in the quasinormal spectrum and stability of the spacetime.

We will perform a perturbative analysis of those spacetimes by computing the quasinormal frequencies due to a probe massless scalar field and in such, study the linear stability of the geometries.

\section{Perturbative analysis: Probe scalar field and quasinormal frequencies}\label{sec.3}

Let us consider the propagation of a probe massless scalar field in the geometry whose dynamical behavior is dictated by the Klein-Gordon equation, 
\be
\label{sc1}
\Box \psi = \frac{1}{\sqrt{-g}}\partial_\mu \left( \sqrt{-g}g^{\mu \nu}\partial_\nu \psi \right) =0.
\ee
In (2+1)-dimensional spherically symmetric spacetimes as that of (\ref{ansatz}), the above equation can be written in the form
\be
\label{sc2}
4 \frac{\partial^2 \Psi}{\partial u \partial v} + V(r) \Psi =0.
\ee
Here $u$ and $v$ are the usual double-null coordinates defined as $du= dt-dr_*$ and $dv= dt+dr_*$ with $dr_* = f^{-1}dr$ representing the tortoise coordinate. The scalar field is decomposed as $\psi (t,r,\varphi)= r^{-1/2}e^{-i \kappa \varphi}\Psi (u,v) $ and the effective potential $V(r)$ is written as
\be
\label{sc3}
V(r) = f \left( \frac{\kappa^2}{r^2}+\frac{f'}{2r}-\frac{f}{4r^2} \right),
\ee
with $\kappa$ being the angular momentum of the field and $'$ denoting a derivative with relation to $r$.

\begin{figure}[h]
\includegraphics[height=5cm, width=8cm]{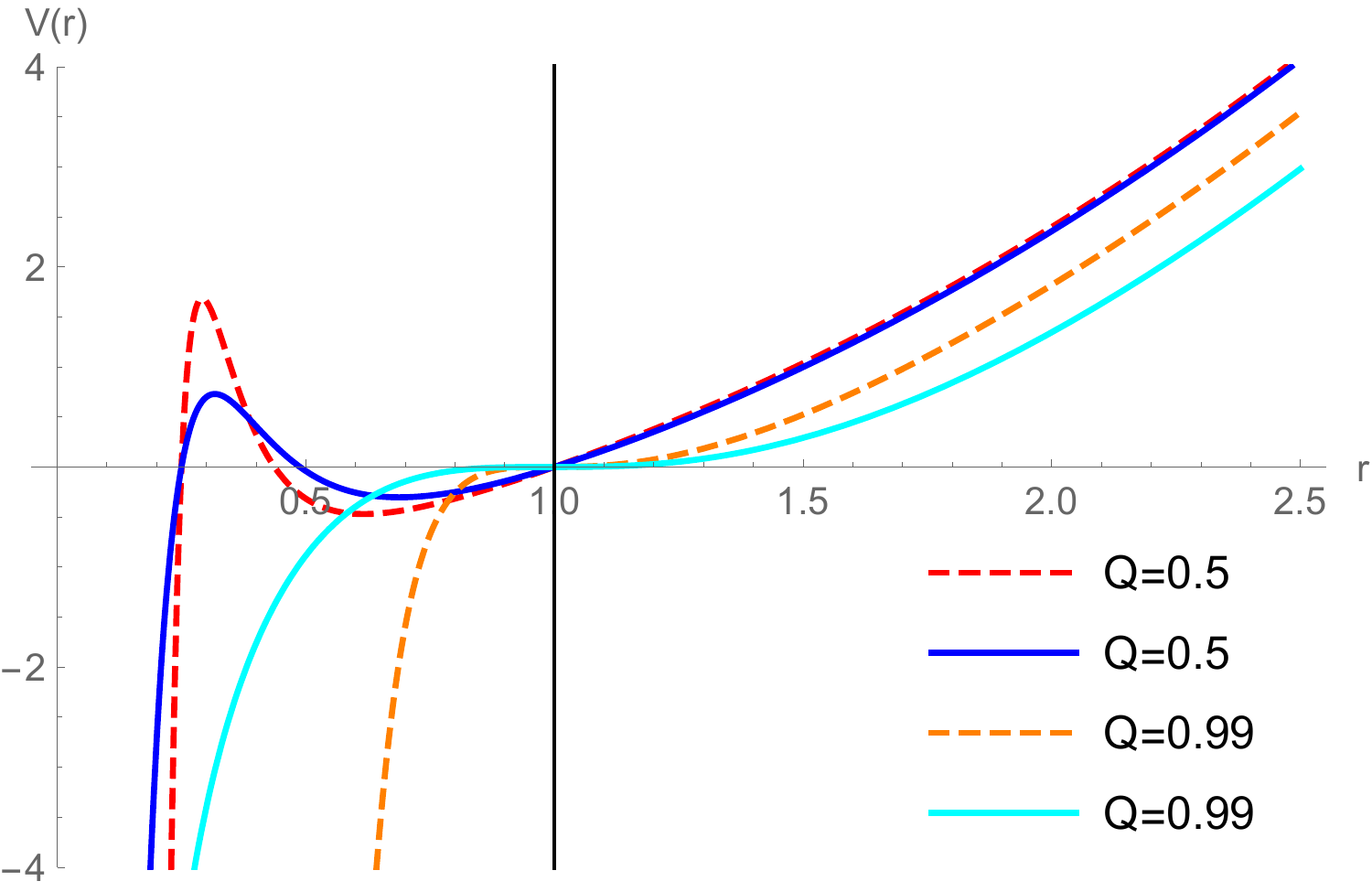}
\includegraphics[height=5cm, width=8cm]{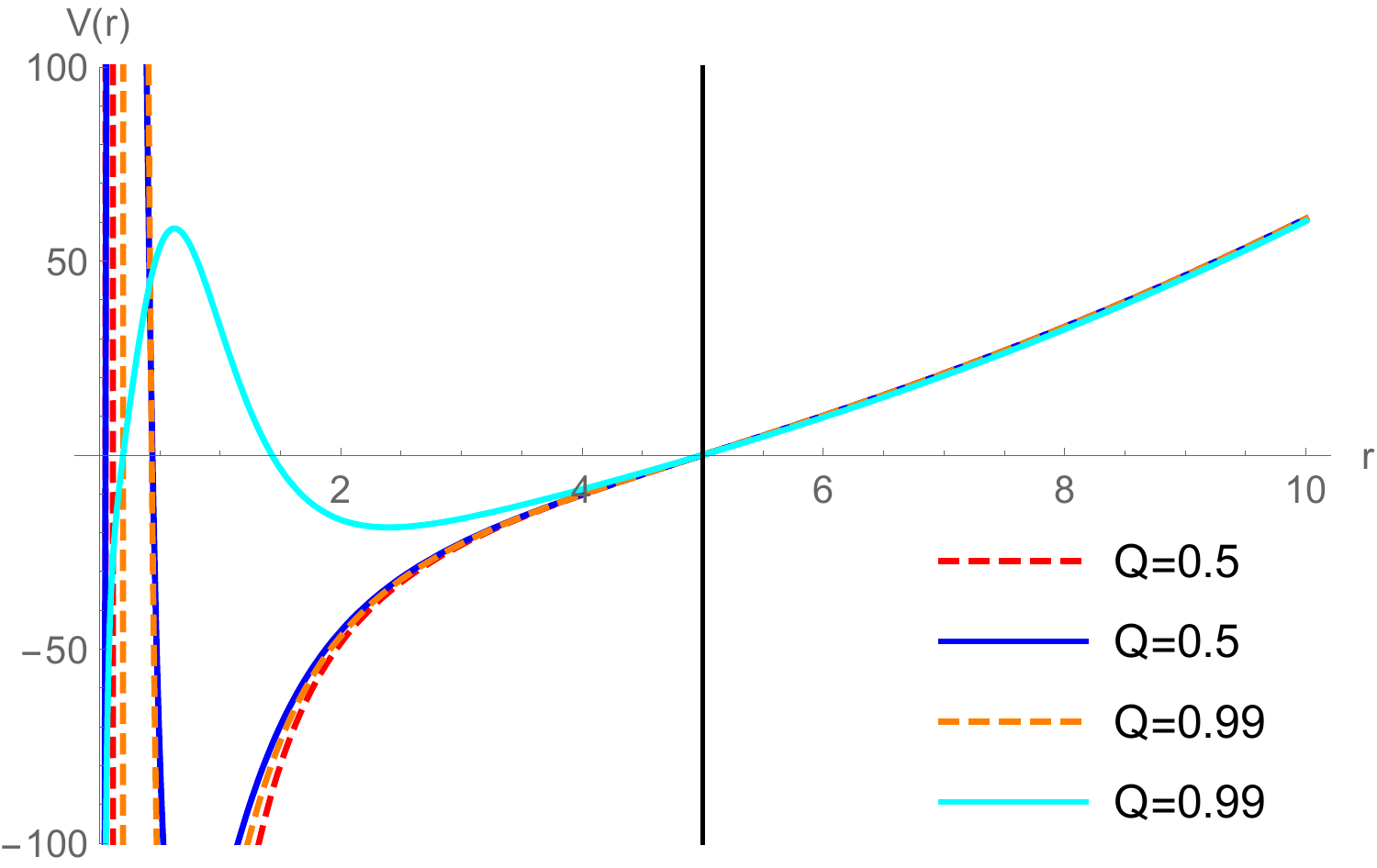}
\caption{Effective potential with two values of charge for singular (dashed) and regular (solid) charged black holes. The parameters used are $L = 1$, $\kappa = 0$ ({\it{left }}) $r_p = 1$, ({\it{right}}) $r_p = 5$. The vertical solid black line represents the event horizon.}
\label{pot_charged_bh}
\end{figure}

\begin{figure}[h]
\includegraphics[height=5cm, width=8cm]{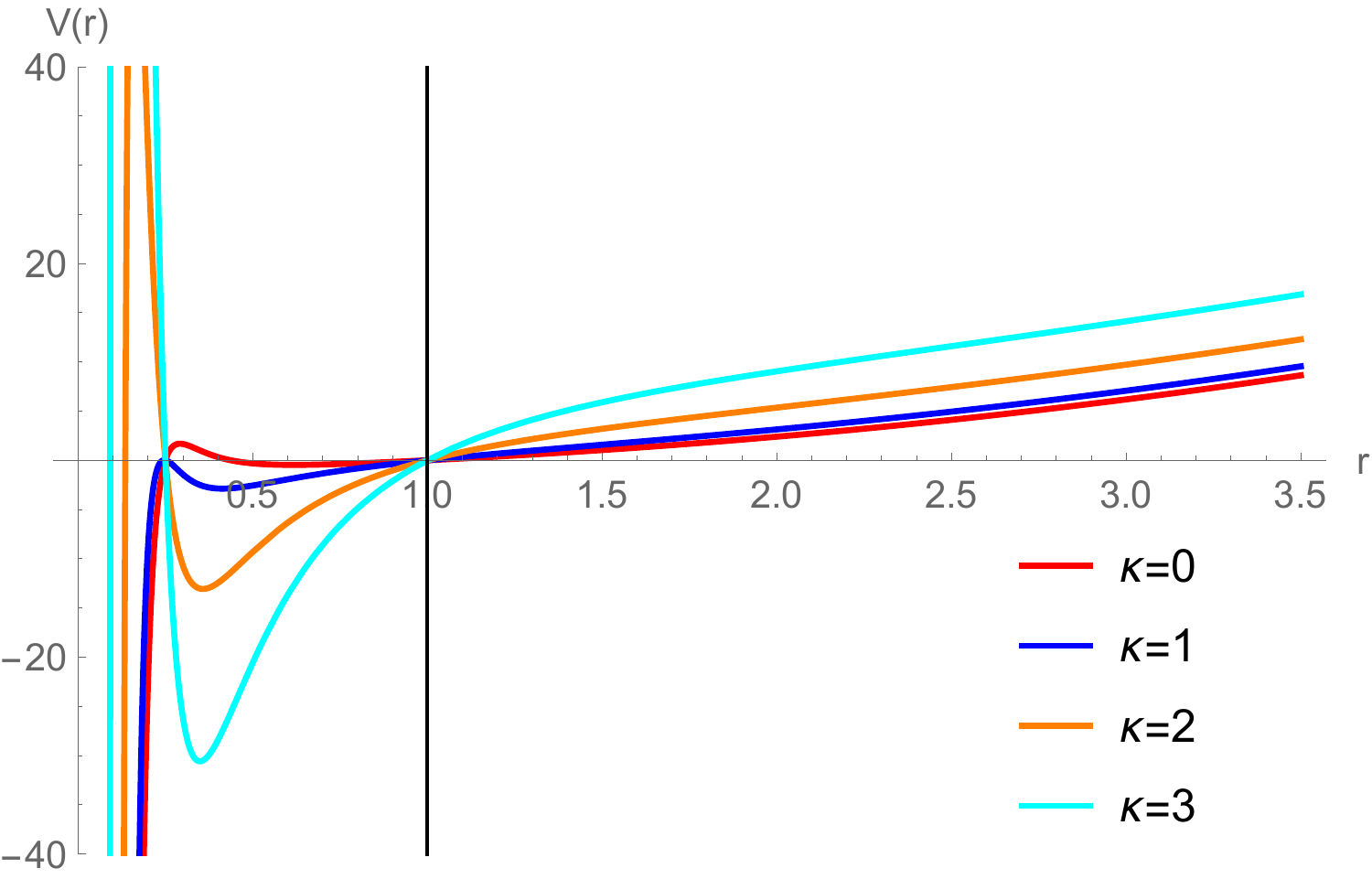}
\includegraphics[height=5cm, width=8cm]{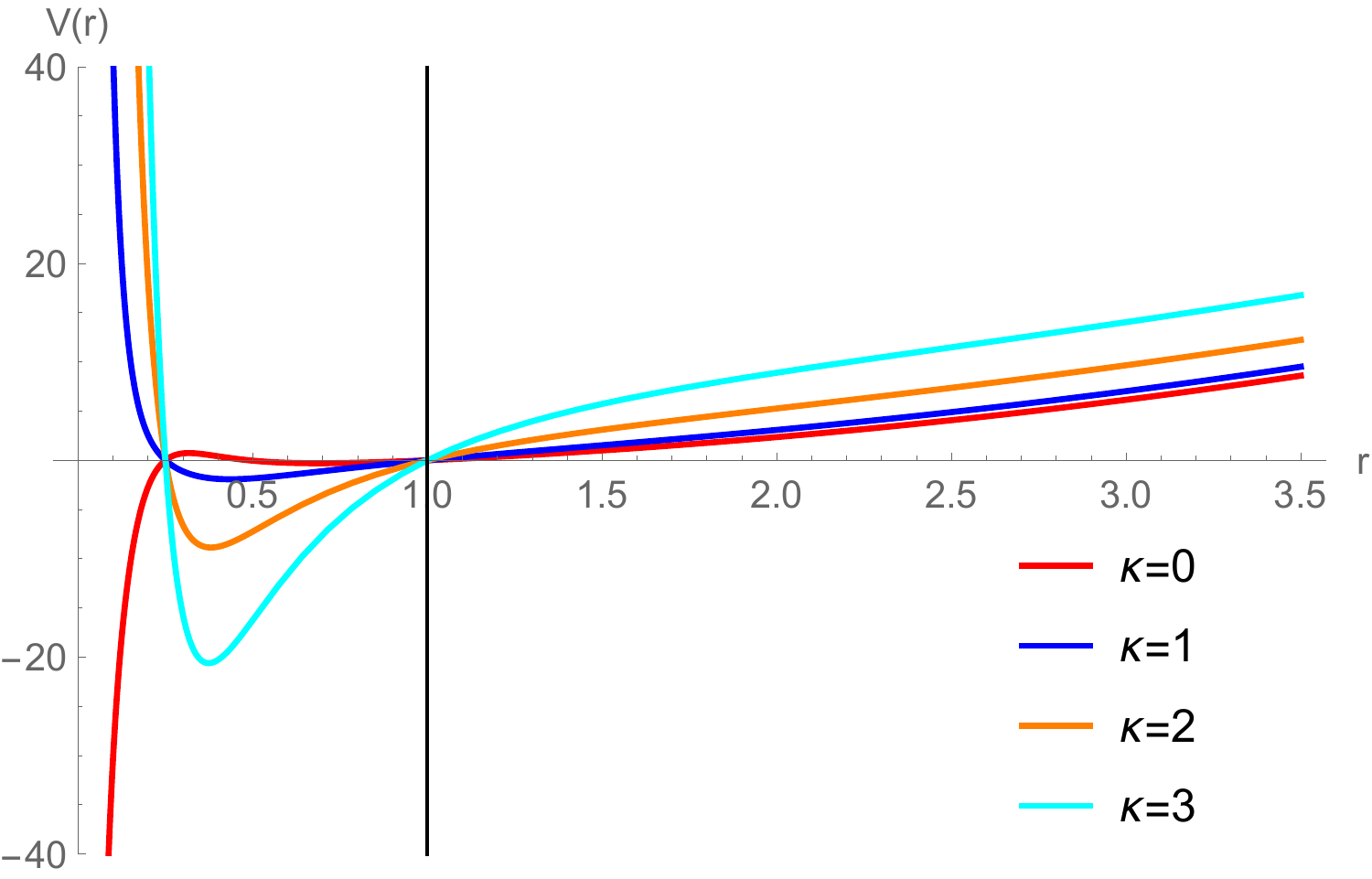}
\caption{Effective potential for singular ({\it{left }}) and regular ({\it{right}}) charged black holes with $L = r_+ = 1$ and different values of angular momentum $\kappa$. The vertical solid black line represents the event horizon.}
\label{pot_charged_bh_kappa}
\end{figure}

\begin{figure}[h]
\includegraphics[height=5cm, width=8cm]{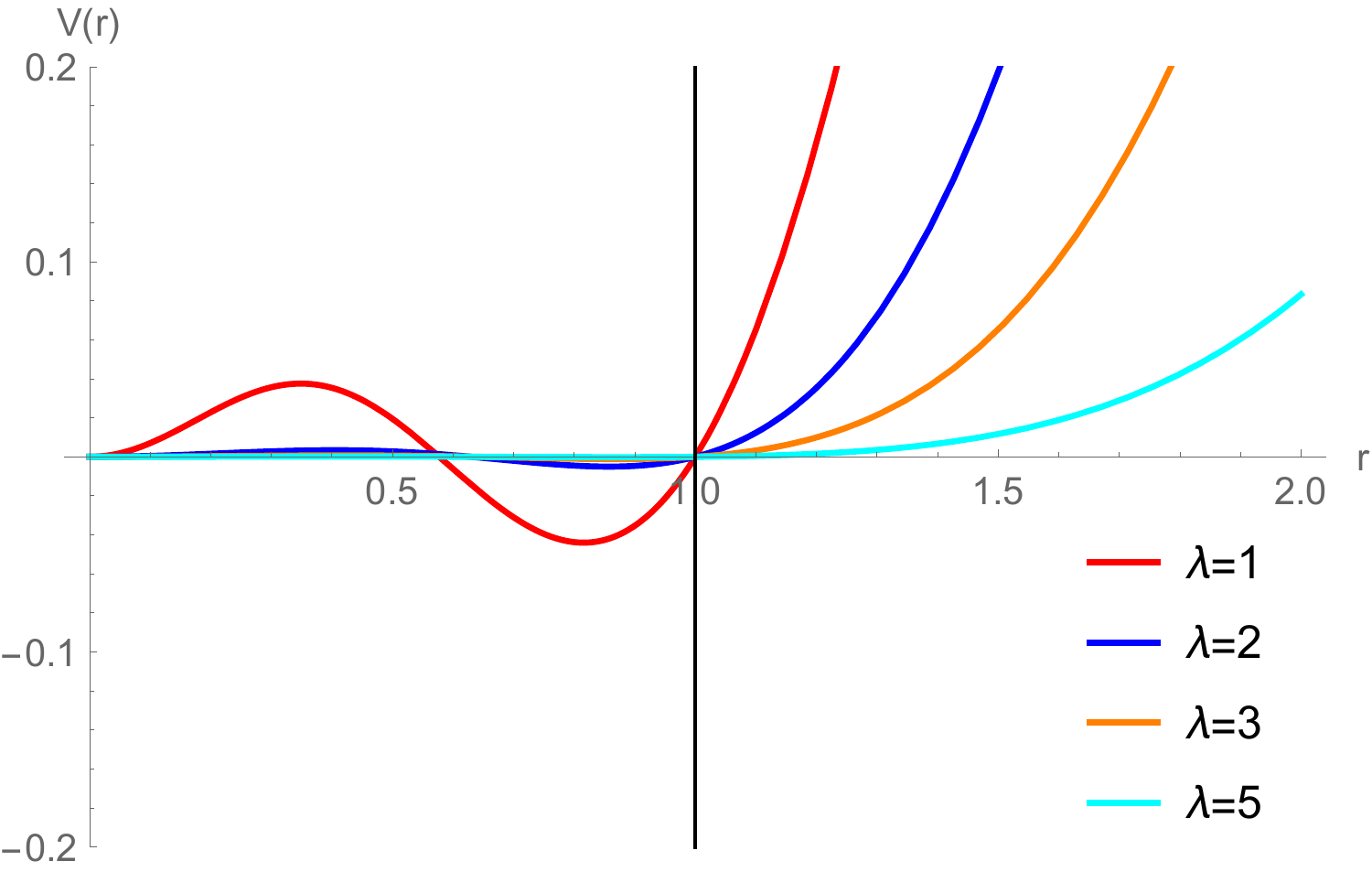}
\includegraphics[height=5cm, width=8cm]{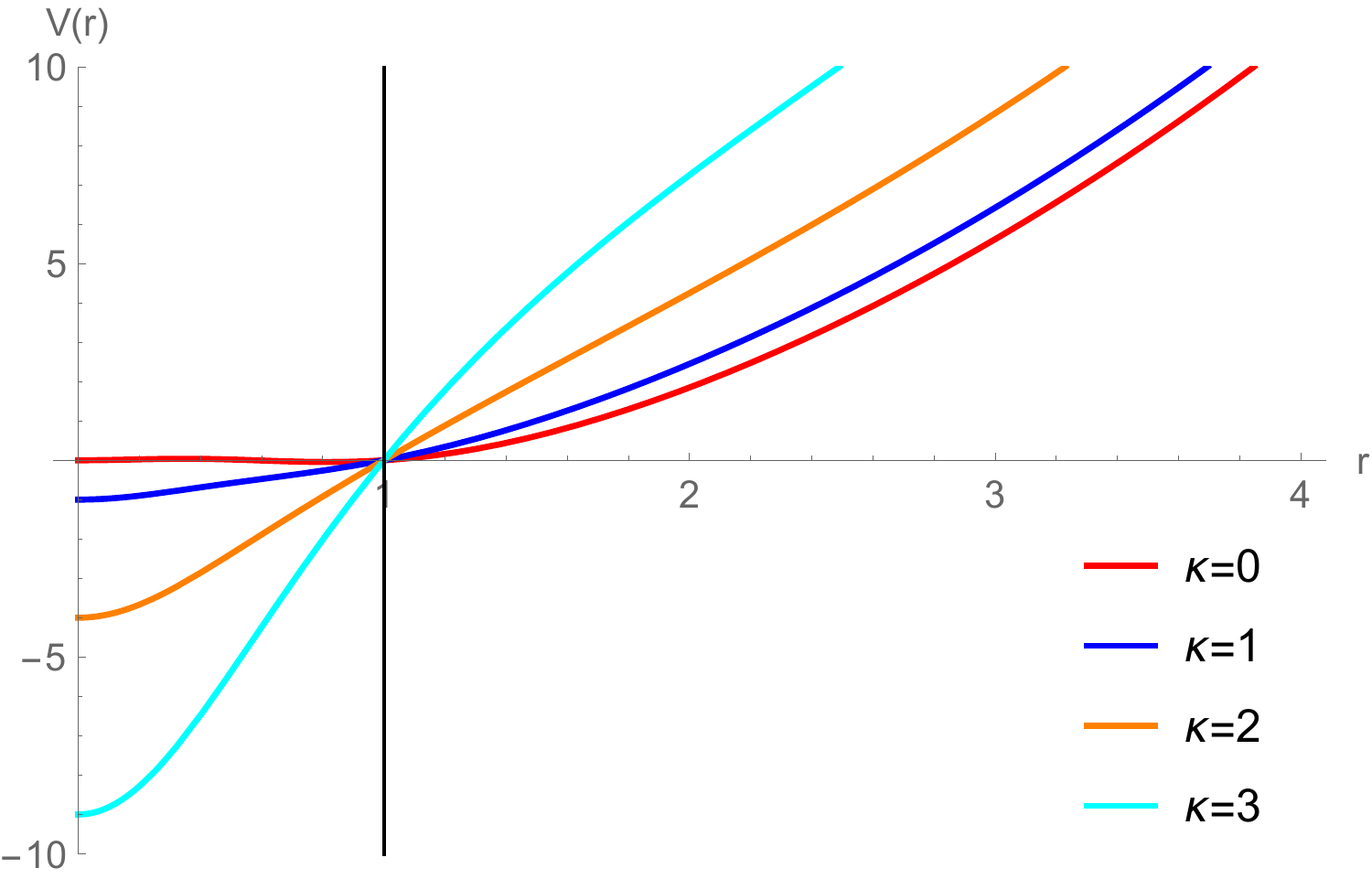}
\caption{Effective potential for regular black hole with $r_p = L = 1$: ({\it{left }}) $\kappa = 0$ and different values of $\lambda$; ({\it{right}}) $\lambda = 1$ and different $\kappa$. The solid black lines represent the event horizons.}
\label{pot_lambda_bh}
\end{figure}

The shape of $V(r)$ dictates the analysis of the dynamical stability of the geometry. Whenever $V(r)>0$ (in the region $r> r_+$) the geometry is proved to be stable to linear perturbations \cite{Horowitz_2000, Gonzalez_2021, Fontana_2019}. On the other hand if $V(r)<0$ at last in some part of $r>r_+$, the geometry can destabilize, depending on ``how negative" $V(r)$ is.
We studied widely the effective potential of all three geometries and obtained $V(r)>0$ whenever $r>r_+$, such that no instabilities are expected in the field evolution, what we confirm with numerical computations. They are given explicitly in the expressions below,
\be
\label{vcsingr}
V_{SBH}=\frac{(r^2-r_+^2) \left(r^2 r_+^2-L^2 Q^4\right) \left[L^2 \left(Q^4 \left(r^2-5 r_+^2\right)+4 \kappa ^2 r^2 r_+^2\right)+r^2 r_+^2 \left(3 r^2+r_+^2\right)\right]}{4 L^4 r^6 r_+^4},\ \ \
\ee
\be
\label{vcregr}
V_{RBH}=\frac{(r^2-r_+^2) \left(r^2 r_+^2-L^2 Q^4\right) \left[L^2 \left(4 \kappa ^2 r_+^2 \left(Q^4+r^2\right)^2-Q^4 \left(Q^4 \left(3 r^2+r_+^2\right)+5 r^2 r_+^2-r^4\right)\right) \right.}{4 L^4 r^2 r_+^4 \left(Q^4+r^2\right)^3}+\ \ \ \nonumber \\
\nonumber\\
+\frac{\left. r^4 r_+^2 \left(7 Q^4+3 r^2\right)+r^2 r_+^4 \left(r^2-3 Q^4\right)\right]}{4 L^4 r^2 r_+^4 \left(Q^4+r^2\right)^3}, \hspace{3cm}
\ee
\be
\label{vlambdar}
V_{\lambda}=\frac{(r^2-r_+^2)\left[4 \kappa ^2 \lambda ^4 L^2+r^4 \left(7 \lambda ^2+4 \kappa ^2 L^2+r_+^2\right)+\lambda ^2 r^2 \left(8 \kappa ^2 L^2-3 r_+^2\right)+3 r^6\right]}{4 L^4 \left(\lambda ^2+r^2\right)^3}.
\ee
Some plots exemplifying the behavior of potentials are given in Figs. (\ref{pot_charged_bh}), (\ref{pot_charged_bh_kappa})and (\ref{pot_lambda_bh}).

From the above expressions we can recognize all effective potentials as similar in the region $r \geq r_+$ presenting significant differences only inside the event horizon $r_+$. We still see, as expected, that the qualitative differences are enhanced by increase in the value of $r_+$. The asymptotic behavior of such effective potentials when $r\to 0$ is given by
\be
\label{vcsingr0}
V_{SBH}\sim -\frac{5 Q^8}{4 r^6}+O\left(r^{-3}\right),
\ee
\be
\label{vcregr0}
V_{RBH}\sim \frac{4 \kappa ^2-1}{4 r^2}+O\left(r^0\right),
\ee
\be
\label{vlambdar0}
V_{\lambda}\sim -\frac{\kappa ^2 r_p^2}{\lambda ^2 L^2}+O\left(r^2\right).
\ee
It is worth to mention that both $V_{SBH}$ and $V_{RBH}$ are singular at the origin, contrary to the one horizon black hole. This fact is particularly important in the regular black hole case, because despite the fact that the hypersurface is regular, the effective potential becomes singular at $r=0$. It can possibly drive a Cauchy horizon instability. When $r\to \infty$ all three potentials go as $V\sim \frac{3 r^2}{4 L^4}$, a characteristic behavior of asymptotically AdS spacetimes.

The fact that $V>0$ in the region $r\geq r_+$ makes possible the investigation of the quasinormal spectra of the black holes taking plane waves as boundary condition (near horizon) with the double null integration technique \cite{Gundlach_1994}. Together with the Prony method \cite{Konoplya:2011qq}, both numerical calculations provide the quasinormal frequencies with good accuracy. As a cross-check method we use the Frobenius expansion similar to that developed in \cite{Horowitz_2000}. In such expansion, we consider the wave equation with the new radial coordinate $z= 1/r$, developing that to
\be
\label{eqe1}
z^4F \frac{\partial^2 \Psi}{\partial z^2} + \big( 2z^3 F - z^2 f' + 2i\omega z^2 \big) \frac{\partial \Psi}{\partial z} + \left(\frac{z^2F - 2z f' - 4\kappa^2 z^2}{4} \right) \Psi = 0
\ee
in which $F= g_{tt}\big|_{r=1/z}$ and  $f'= (\partial_r g_{tt})\big|_{r=1/z}$. The Klein-Gordon field is now adjusted to the ansatz $\psi = e^{-i\omega t + i\kappa \varphi} \Psi (z)$. The solution of the equation considers a series expansion around the event horizon $z_+ = r_+^{-1}$ and the first boundary condition for quasinormal modes ingoing plane waves at $z_+$ implicitly implemented in the series expansion, $\Psi \rightarrow \sum_n(z-z_+)^{n+\alpha }$. That corresponds to the choice $\alpha = 0$, in accordance with (\ref{eqe1}). Finally the method consists in truncating the series in a certain number of terms with the quasinormal equation $\Psi (0)=0$ numerical implemented, which represents the second boundary condition, typical of AdS spacetimes.

\subsection{Singular and regular black holes with charge}

The singular black hole with charge represented by the line element (\ref{bh_singular})  possesses two regular horizons. The most external of those, the event horizon characterizes the boundary of the scattering problem and in that region, the plane wave condition reads
\be
\label{}
\Psi |_{r_* \rightarrow -\infty} \rightarrow e^{-i\omega r_*}
\ee
 Since the scalar potential diverges in the AdS infinity ($r_*=0$ or $r=\infty)$, besides the usual plane wave condition at the horizon, we require that
\be
\label{sc4}
\Psi |_{r_*=0}\rightarrow 0
\ee
which settles the scattering question down.

The fundamental quasinormal modes obtained for the singular and regular charged black holes are displayed in Table \ref{tb1}.

\begin{table}[h]
  \centering
 \caption{{\color{black} Quasinormal modes of the massless scalar field with $\kappa=0$. For simplicity we use a fraction of the charge, $Q= \sqrt{r_+r_-/L}$ relative to its maximum value, $Q_{max}=r_+$ defined as $\mathcal{R}= Q/Q_{max}$. All frequencies are purely imaginary and stable, $\Im (\omega ) < 0$}.}
\addtolength\tabcolsep{6pt}
    {\color{black}\begin{tabular}{c|ccc}
    \hline    \hline
Singular & $r_+=1$& $r_+=10$& $r_+=100$\\
    \hline
$\mathcal{R}$	 &	\multicolumn{3}{c}{$\Im (\omega )$}\\
\hline \hline
0	&	-1.9996	&-19.931&	-199.306 	\\
0.1	&	-1.9928	&-19.835&	-198.075\\
0.2	&	-1.9320	&-19.200&	-192.001\\
0.3	&	-1.8292	&-18.200&	-182.001\\
0.4	&	-1.6907	&-16.800&	-168.001\\
0.5	&	-1.5136	&-15.000&	-150.002\\
0.6	&	-1.2814	&-12.800&	-128.000\\
0.7	&	-1.0208	&-10.201&	-102.000\\
0.8	&	-0.7203	&-7.2002&	-72.000	\\
0.9	&	-0.3804	&-3.8000&	-38.000	\\
    \end{tabular}}
    \addtolength\tabcolsep{3pt}
    {\color{black}\begin{tabular}{c|ccc}
    \hline    \hline
Regular &$r_+=1$&$r_+=10$&$r_+=100$ \\
\hline
$\mathcal{R}$	 &	\multicolumn{3}{c}{$\Im (\omega )$}\\
\hline \hline
0	&	-1.9996	      &-19.927	     &-199.182  \\
0.1	&	-1.9244	 	  &-18.007	     &-112.998  \\
0.2	&	-1.7885	 	  &-13.999	     &-96.272	\\
0.3	&	-1.6132	 	  &-10.677	     &-91.012	\\
0.4	&	-1.3866	 	  &-8.8256		 &-84.001\\
0.5	&	-1.1509	 	  &-7.6029		 &-75.000\\
0.6	&	-0.9061	 	  &-6.4244		 &-64.000\\
0.7	&	-0.6625	      &-5.1057	     &-51.000\\
0.8	&	-0.4283	      &-3.6012	     &-36.000\\
0.9	&	-0.2772	      &-1.9001	     &-19.000\\
    \end{tabular}}

  \label{tb1}
\end{table}
As expected, these modes are purely imaginary and under specific condition have a worth scaling with the black hole parameters written as
\be
\label{sc6}
\omega_{sing} = -2ir_+(1-\mathcal{R}^2) = -2i(r_+-r_-),
\ee
\be
\label{sc7}
\omega_{reg} = - ir_+(1-\mathcal{R}^2)= - i(r_+-r_-),
\ee
for singular and regular charged black holes respectively. The fundamental modes (singular geometry) expressed by (\ref{sc6}) follow very strictly the same scaling calculated in \cite{0101194} (left branch) which is strictly related to the critical phenomena of the black hole formation.
 
In the singular spacetime the variance of (\ref{sc6}) to the data of Table \ref{tb1} for larger black holes is smaller than the usual deviation using different methods to quasinormal modes (in some cases to the 6th figure). In the pure BTZ limit ($r_-=0$), our results with double-null integration presented here are about $0.2\%$ to $0.3\%$ deviant from the analytical frequencies \cite{Cardoso_2001}. Relation (\ref{sc6}) couples the results of Table \ref{tb1} almost perfectly ($r_+>1$) and recovers the pure BTZ case whenever $r_- =0$ \cite{Cardoso_2001, de_Oliveira_2018}.

\begin{figure}[h]
\includegraphics[scale=0.5]{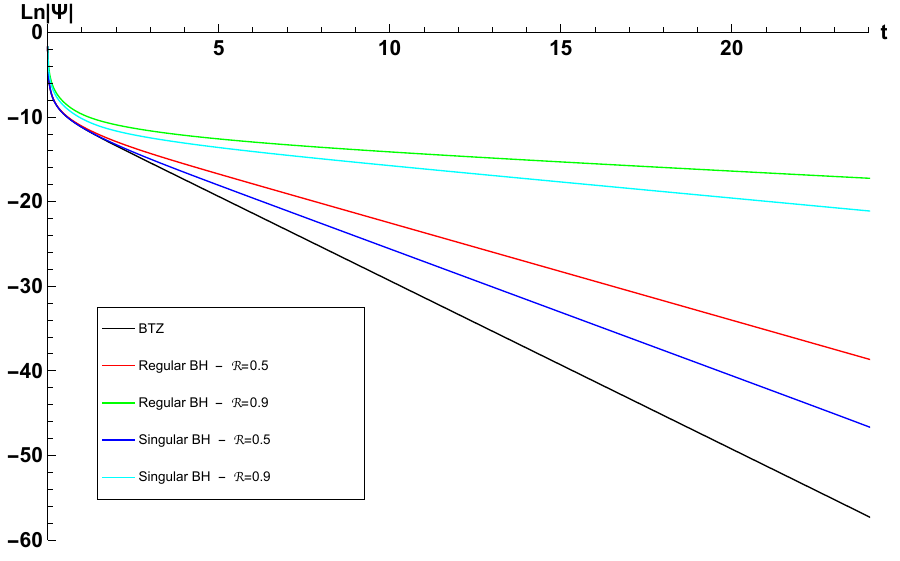}
\includegraphics[scale=0.5]{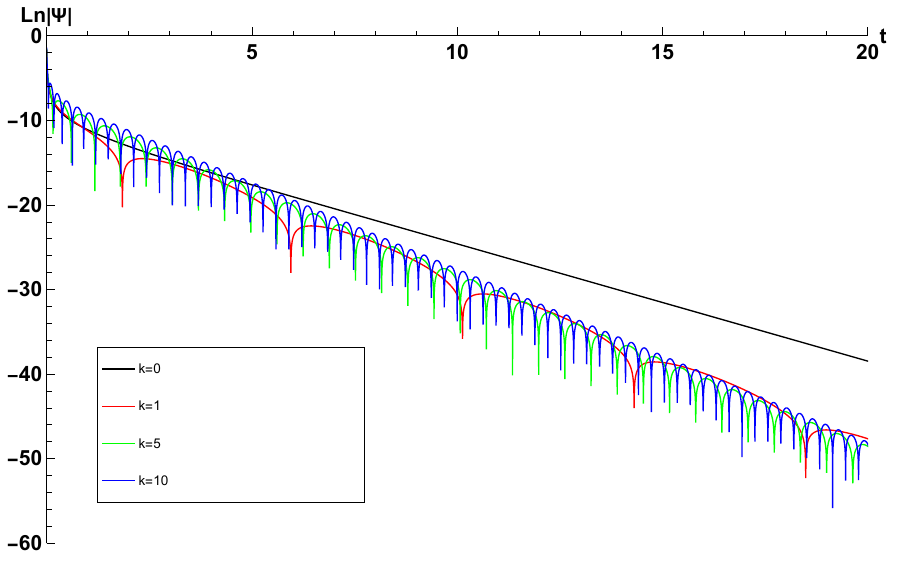}
\caption{Scalar field profiles of charged black holes in (2+1) dimensions. In the left panels purely imaginary evolutions are shown for singular/regular black holes with $\kappa=0$. In the right panels we can see field evolutions with different angular momentum in a regular black hole with $\mathcal{R}=0.4$ and $r_+=1$.}
\label{fd1}
\end{figure}

In the charged regular geometry, contrarily to what is found in typical BTZ-like black holes with spherical symmetry the frequencies are not a linear function of size of the hole when $r_+$ is small. In regimes of intermediate to high spacetime charge and hole size ($r_+$) such scale is recovered and the expression (\ref{sc7}) arises representing half the value of the scaling of the singular case. We show the typical scalar field evolution in Fig, \ref{fd1}.
The attenuation of $\omega$ with increasing black hole charge is typical for charged BTZ-like black holes \cite{Cuadros_Melgar_2022,f23quasi}. In fact, in the purely charged BTZ geometries, such weakening is even more pronounced: for small black hole charges, the damping of the fundamental mode lowers ca. $30\%$ when the black hole charge diminishes $10\%$. In our case, a smaller depletion is observed as a result of the electromagnetic potential shape which is similar to the $(3+1)$ geometries (and very different from the charged BTZ). In this case, the conditions for superradiance and unstable mode evolutions for a charged scalar field will be affected, generating a specific spectrum very different from the pure BTZ geometry \cite{f23inst} to be addressed in separate lines of investigation for another works.

\subsection{Regular black hole with one horizon}

The last black hole we analyze is the regular BTZ solution with a three parameter lapse function as that of (\ref{func_horizonte}). In such case the extra parameter $\lambda$, acts similarly to an AdS cosmological term, attenuating the value of $\Im (\omega )$ as it increases. A table with  quasinormal spectra and different $\lambda$ is presented in Table \ref{tb3}.
\begin{table}[h]
  \centering
 \caption{{\color{black} Quasinormal modes of the massless scalar field with $L=1$ and $\kappa=0$. All frequencies are purely imaginary and stable, $\Im (\omega ) < 0$}.}
\addtolength\tabcolsep{6pt}
    {\color{black}\begin{tabular}{c|cccccc}
    \hline    \hline
$\lambda$ & $10^{-2}$ & $10^{-1}$ & $0.5$ & 1 & 5 & 10 \\
    \hline
 & \multicolumn{6}{c}{$\Im (\omega )$} \\
$r_+ = 1$  & -1.98019 & -1.8019 & -1.1259 & -0.6171 & -0.04319 & -0.01107 \\

$r_+ = 10$ & -19.9820 & -19.8019 & -19.0036 & -18.0178 & -11.2586 & -6.17090 \\
    \hline    \hline
    \end{tabular}}
  \label{tb3}
\end{table}

The values of Table \ref{tb3} were obtained with the Frobenius method and confirmed with good accuracy with the characteristic integration scheme. Their deviation is no higher than $0.1\%$ (except for the highest $\lambda$ case with a difference of about $1\%$).

\begin{table*}
  \centering
 \caption{{\color{black} Quasinormal modes of the massless scalar field with $r_+=L=1$. All frequencies are stable, $\Im (\omega ) < 0$}.}
\addtolength\tabcolsep{3pt}
{\color{black}\begin{tabular}{c|cccccccc}
\hline 
\hline
$(\kappa=1)$, $\lambda$   & $0$     & $0.1$   & $0.2$    & $0.3$   & $0.4$   &$0.5$      & $1$        & $2$ \\
\hline
$\Im ( \omega )$      & -2.0001 & -2.0051 & -2.0188  & -2.0377 & -2.0507 & -1.9891   & -0.8774   & -0.4011\\
$\Re (\omega)$        & 1.0002 & 0.9750 & 0.8971  & 0.7578 &0.5309  & 0         & 0         & 0\\
\hline
\hline
$(\kappa=2)$, $\lambda$   & $0$      & $0.1$    & $0.2$     & $0.3$    & $0.4$   & $0.5$     & $0.6$      & $0.8$ \\
\hline
$\Im (\omega)$        & -2.00028 &-2.00810  & -2.03158  & -2.06926 &-2.11890  & -2.17735 & -2.24025   & -2.24499\\
$\Re ( \omega )$      & 2.00077 & 1.98632 &1.94430   & 1.87412 & 1.77521 & 1.64605 & 1.48336   & 0\\
\hline
\hline
$(\kappa=3)$, $\lambda$    & $0$      & $0.1$   & $0.2$     & $0.3$    & $0.4$   & $0.5$    & $0.6$    & $0.9$ \\
\hline
$\Im (\omega)$        & -2.00013 &-2.00949 & -2.03557  & -2.07833 &-2.13548 & -2.20480 & -2.28320 & -2.55336\\
$\Re ( \omega )$      & 3.00070 & 2.99032 &2.96226  & 2.91406 & 2.84737 & 2.76218 & 2.65747 & -2.21252\\
\hline 
\hline
    \end{tabular}}
  \label{tb3a1}
\end{table*}


\begin{table}[h]
 \centering
 \caption{{\color{black} Quasinormal modes of the massless scalar field with $r_+=10,L=1$. All frequencies are stable, $\Im (\omega ) < 0$}.}
\addtolength\tabcolsep{3pt}
{\color{black}\begin{tabular}{c|cccccccc}
\hline   
\hline
$(\kappa=1)$, $\lambda$      & $0$      & $0.1$     & $0.4$    & $1.0$   & $5.0$     & $10.0$     & $15.0$        & $20.0$ \\
\hline
$\Im ( \omega )$        & -20.0015 & -20.0015  & -20.0016 & -18.2839 & -11.3060 & -6.19507   & -3.63687     & -2.31929\\
$\Re (\omega)$          & 1.00020 & 0.979889 & 0.598966& 0        &  0       & 0          & 0            &     0   \\
\hline    
\hline
$(\kappa=2)$ , $\lambda$     & $0$      & $0.1$     &  $0.5$     & $0.9$     & $5.0$     & $10.0$     & $15.0$      & $20.0$ \\
\hline
$\Im (\omega)$          & -20.0023 & -20.0023  & -20.0030   & -20.0029  & -11.4496  & -6.26780  & -3.69071   & -2.36564\\
$\Re ( \omega )$        & -2.00047 & -1.99025  & -1.72818   & -0.864540 &     0     &     0     &    0       &     0   \\
\hline    
\hline
$(\kappa=3)$, $\lambda$      & $0$      & $0.1$     & $0.5$     & $1.0$    & $1.4$    & $2.0$     & $3.0$      & $5.0$ \\
\hline
$\Im (\omega)$          & -20.0023 & -20.0023  & -20.0041  & -20.0072 & -20.0047  & -17.4402 & -15.1172   & -11.6935\\
$\Re ( \omega )$        & -3.00071 & -2.99376  & -2.82250  & -2.21500 & -2.84737  &    0     &     0      &    0    \\
\hline    
\hline
    \end{tabular}}
  \label{tb3a10}
\end{table}

In the Tables \ref{tb3a1} and \ref{tb3a10} we list the calculated frequencies with angular momentum obtaining an interesting feature in that spectra: the quasinormal modes change from an oscillatory pattern to a purely imaginary character as we increase $\lambda$. Such change may be related to the presence of a secondary family of modes taking control of the field evolution. Interestingly we can see that $\Im (\omega )$ goes in different directions with increasing $\lambda$. In the primary family, the increase in $\lambda$ rises the damping of the oscillation and in the purely imaginary scenario, the opposite happens. The existence of multiple families of oscillations were described in many references along the last years \cite{Cardoso_2018,Destounis_2019,Fontana:2022whx} though its presence has to be verified with analytical methods identifying overtone numbers for each family.

\section{Final remarks}\label{sec.final}

Black hole solutions in lower dimensions are a fruitful scenario to test ideas and proprieties of general relativity solutions in $(3+1)$ dimensions. This paper analyzed three representative families of (2+1)-black hole solutions in the EM-GQT theories, with two of them featuring Cauchy and event horizons with and without curvature singularity at the origin. Those black holes share almost the same causal proprieties as in the Reissner-Nordstr\"om-AdS black hole in $(3+1)$ dimensions. In the case of the singular solution, the maximal extension is the same, and for the regular solution, the timelike singularity at $r=0$ is absent, as expected. After crossing the Cauchy horizon, an uncharged massive particle can avoid the timelike singularity due to a repulsive gravitational force near $r=0$, as in the Reissner-Nordst\"om solution. The same occurs in the case of the regular solution, but only for particles with angular momentum $k\neq 0$.

The third black hole family is regular with one horizon and recovering the BTZ solution when the parameter $\lambda = \beta_{0} L^{2} p^{2}$ goes to zero. The causal structure shows that a timelike particle will cross the well-behaved surface $r=0$ and emerge asymptotically with an AdS white hole as in the diagram on the right side of Fig. (\ref{diagrams}).

We tested all three families of black holes against a scalar field introduced in the geometry as 
a probe to the background geometry. To linear order, we can see the scalar field decay in time in sets of quasinormal modes with a very peculiar structure. In the singular charged black hole, we observe the fundamental mode to scale the distance between horizons just as proposed in the rotating case \cite{0101194}, which is an astonishing novelty since the geometry, contrary to that of \cite{0101194} is static. Another interesting scaling arrives in the regular black hole for intermediate and near extremal charges, proportional to the same distance but with an attenuation of 1/2.

The regular black hole with only one horizon possesses another interesting quasinormal mode spectral structure denoting the presence of oscillatory families competing with purely imaginary frequencies to dominate the field profile. Depending on the black hole parameters, one or another cast as the most resilient profile.

Further lines of investigation in view of quasitopological black holes we study here are the test of different probe fields in such geometries and instabilities delivered by a charged scalar perturbation as usual of charged geometry, which can be seen in the simplest charged BTZ black hole \cite{f23inst}.

\begin{acknowledgments}
We thank Pablo Bueno for useful discussions. This work was partially supported by CNPq (Conselho Nacional de Desenvolvimento Científico - Brazil) under Grant No. 405749/2023-6. R. D. B. F is supported by the Center for Research and Development in Mathematics and Applications (CIDMA) through the Portuguese Foundation for Science and Technology (FCT-Fundação para Ciência e a Tecnologia), references https://doi.org/10.54499/UIDB/04106/2020 and https://doi.org/10.54499/UIDP/04106/2020.
\end{acknowledgments}

\section*{References}

\bibliography{references}

\end{document}